\documentclass[%
 reprint,
superscriptaddress,
amsmath,amssymb,
aps,
pre,
showpacs,
longbibliography
]{revtex4-1}

\usepackage[utf8]{inputenc}
\usepackage{graphicx}
\usepackage{dcolumn}
\usepackage{bm}
\usepackage{tikz}
\usepackage{tabularx}

\usepackage{amssymb}
\usepackage{pgfplots}
\pgfplotsset{compat=1.3}
\usetikzlibrary{plotmarks}
\usepackage{rotating}
\renewcommand{\th}[1]{\textcolor{black}{#1}}

\newcommand{\usv}{^{\rm loc}}

\newcommand{\rheo}{}

\begin{document}
\title{Predicting and assessing rupture in protein gels under oscillatory shear}

\author{Brice Saint-Michel}
 \email{brice.saint-michel@ens-lyon.fr}
\author{Thomas Gibaud}
 \email{Corresponding author, thomas.gibaud@ens-lyon.fr}
 \affiliation{Univ Lyon, Ens de Lyon, Univ Claude Bernard, CNRS, Laboratoire de Physique, F-69342 Lyon, France}%

\author{S\'ebastien Manneville}
	\email{Corresponding author, sebastien.manneville@ens-lyon.fr}
 \affiliation{Univ Lyon, Ens de Lyon, Univ Claude Bernard, CNRS, Laboratoire de Physique, F-69342 Lyon, France}%

\date{\today}

\begin{abstract}
Soft materials may break irreversibly upon applying sufficiently large shear oscillations, a process which physical mechanism remains largely elusive. In this work, the rupture of protein gels made of sodium caseinate under an oscillatory stress is shown to occur in an abrupt, brittle-like manner. Upon increasing the stress amplitude, the build-up of harmonic modes in the strain response can be rescaled for all gel concentrations. This rescaling yields an empirical criterion to predict the rupture point way before the samples are significantly damaged. ``Fatigue'' experiments under stress oscillations of constant amplitude can be mapped onto the former results, which indicates that rupture is independent of the temporal pathway in which strain and damage accumulate. Finally, using ultrasonic imaging, we measure the local mechanical properties of the gels before, during and after breakdown, showing that the strain field remains perfectly homogeneous up to rupture but suddenly gives way to a solid--fluid phase separation upon breakdown.
\end{abstract}

                             
\maketitle

\section{Introduction}
Under sufficiently large stresses, solid materials may either flow or break. In hard solids, such as metals or composite structures, these highly nonlinear phenomena have a century of history for an obvious reason: in an engineering context, it is critical to understand failure mechanisms in order to improve mechanical performance and ensure a safe usage. For soft solids, like gels or pastes, the topic is more recent yet nonetheless important. Most soft materials, including colloidal gels~\cite{Coussot2002, Moller2006,Gibaud2008, Gibaud2009, Gibaud2010, Divoux2012,Perge2014, Gibaud2016}, colloidal glasses~\cite{Besseling2007}, foams~\cite{Gilbreth2006,Kabla2007}, star polymers~\cite{Rogers2008} and emulsions~\cite{Becu2006,Goyon2008,Knowlton2014}, display a {\it yielding} transition where they evolve from a solid-like state to a fluid-like state as the stress is increased \cite{Balmforth:2014,Bonn:2015}. This transition is reversible and the material regains its elasticity once brought back to rest. Yielding has tremendous implications for food and cosmetics, including industrial processing and texturing~\cite{Szczesniak2002, Fischer2011, Gibaud2012}. It is also relevant to biology, for instance to the mechanosensing properties of cellular tissues~\cite{Guevorkian2011}. However, when strong interparticle bonds are involved, soft solids rather develop fractures and \textit{irreversibly break} under stress. This is the case for a number of biopolymer gels, such as gelatin~\cite{Baumberger2006}, agar~\cite{Bossi2016} and casein gels~\cite{Leocmach2014}. 

Experimental characterization of the yielding transition is not always easy for soft solids, due to slippery boundary conditions~\cite{Lettinga2009, Meeker2004, Gibaud2009, Ballesta2012, Seth2012,Divoux2016}, long transient regimes or spatial heterogeneities~\cite{Gibaud2009,Perge2014,Gibaud2016, Divoux2010}. Predicting failure in soft solids is even more difficult and remains an open problem. The onset of yielding and/or rupture under a stress load has been associated with a diverging correlation length scale~\cite{Goyon2008}, plasticity~\cite{Lin2014}, ``hot spots''~\cite{Amon2012, Lebouil2014}, and complex avalanche-like statistics~\cite{Dahmen2011}. Nevertheless, such features are insufficient to fully account for the irreversible rupture of soft solids, \th{in particular to predict the failure time, as the process generally involves crack nucleation and fracture growth through the local concentration of stress or strain.} 

\th{A large number of studies have been devoted to cracks and fractures induced by a {\it constant} load or strain in various soft materials including gels~\cite{Bonn1998,Baumberger2006,Leocmach2014}, fibrous materials such as fiberglass and wood \cite{Guarino2002} or paper \cite{Salminen:2002,Rosti2010,Koivisto2016} and harder composite materials \cite{Nechad2005} or even basalt \cite{Heap2011}. Some features, such as the power-law creep that precedes catastrophic failure \cite{Miguel2002,Leocmach2014}, are strikingly reminiscent of brittle failure in harder materials and are well captured by fiber-bundle models \cite{Nechad2005,Pradhan2010,Halasz2012}. Surprisingly,} the irreversible rupture of soft solids under {\it oscillatory} shear remains much less explored, although oscillations provide a convenient way to follow the build-up of nonlinearity as a function of load amplitude, frequency and time~\cite{Gibaud2010,Perge2014,Brader2010,Pouzot2006,Vliet1995,Bremer1990}. Oscillatory loads are also frequently involved in applications and often involve damage accumulation prior to rupture, a phenomenon known as ``fatigue" that is still poorly understood~\cite{Suresh:1998,Kun2007,Kun2008,Miksic2011,Gibaud2016}. 

In this paper, we focus on protein gels made of sodium caseinate under oscillatory stress. Such gels were reported to display brittle-like rupture both under constant stress or strain~\cite{vanVliet:1996,Leocmach2014,Keshavarz:2016}. Here, we show through a Fourier analysis of the strain response that their nonlinear mechanical properties can be scaled on a single master curve independently of the gel concentration.  An empirical criterion is then introduced to predict the failure point based solely on the nonlinear viscoelastic deformation regime where the material remains essentially undamaged. We further show that ``fatigue'' experiments, where damage accumulation leads to rupture, can be also mapped onto the previous results by eliminating the time dependence, suggesting that the rupture process is independent of the specific temporal pathway. Using a recently-developed echography technique, we measure the local viscoelastic properties of the system and find that they remain homogeneous up to rupture, at least on length scales larger than 50~$\mu$m. We finally visualize the rupture process and evidence a phase separation of the system into a fluid region and a solid region after breakdown. 

\section{Materials and methods}
\label{sec:mm}

\subsection{Oscillatory shear experiments}
\label{sec:protocol}

Oscillatory shear is applied thanks to a stress-controlled rheometer (TA Instruments AR G2) equipped with a concentric-cylinder (or Taylor--Couette) geometry with smooth, polymethylmethacrylate (PMMA) walls. The inner rotating cylinder has a radius $R_{\rm i} = 23$~mm, a height $H = 60$~mm and its bottom is terminated by truncated cone of angle $2^\circ$. The fixed outer cylinder has a radius $R_{\rm o}= 25$~mm, leaving a radial gap $e = 2$~mm between the two cylinders (see sketch in Fig.~\ref{fig:prise_fsweep}). The temperature is controlled by a water circulation around the Taylor-Couette cell and set to $25\pm 0.1^\circ$C for all experiments. The samples are gelled in-situ as explained below in Sect.~\ref{sec:gel} and subsequently submitted to an imposed oscillatory shear stress $\sigma\rheo(t) = \sigma_1 \cos(2 \pi f t)$ with frequency $f$ and amplitude $\sigma_1$. The sample strain response $\gamma\rheo(t)$ is recorded by the rheometer. Under medium and large amplitude oscillatory stresses, $\gamma\rheo(t)$ becomes nonlinear~\cite{Hyun2011}. This results in the presence of harmonics in the Fourier series decomposition of $\gamma\rheo(t)$:
\begin{equation}
\gamma\rheo(t) = \sum_k \gamma_k\rheo \cos \left ( 2 \pi k f t + \phi_k\rheo \right ) \,,
\label{eq:gamma_rheo_def2}
\end{equation}
where $\gamma_k\rheo$ is the amplitude of the $k^{\rm th}$ harmonics and $\phi_k\rheo$ its phase with respect to $\sigma\rheo(t)$. In our experiments, even harmonics are always negligible as expected from symmetry considerations in the absence of wall slip~\cite{Graham1995,Reimers1996,Klein2007,Hyun2011,Guo2011,Saintmichel2016} so that we only focus on odd harmonics $k=3$, 5, 7 and 9.  

Two types of oscillatory shear experiments are performed at $f=0.1$ or 0.3~Hz: (i)~``instantaneous'' experiments where the gel is solicited over a short duration and (ii)~``fatigue'' experiments where oscillations are applied with a given stress amplitude for much longer durations. \th{We use those two sets of experiments to predict and probe rupture}. On the one hand, ``instantaneous'' experiments consist of \th{10 oscillations, which represents a good compromise between accuracy on the Fourier coefficients and minimization of damage due to fatigue at large stress amplitudes. This gives us an quasi-instantaneous snapshot of the gel mechanical properties}. Between two such experiments, the gel is left to recover at rest for 10~min. On the other hand, ``fatigue'' experiments allow us to monitor the evolution of the material as it accumulates damage over time. In this latter case, a fresh sample is required for each experiment. 

\subsection{Local Oscillatory Rheology from Echography}
\label{sec:lore}

Local Oscillatory Rheology from Echography (LORE) synchronizes oscillatory shear experiments with ultrafast ultrasonic echography~\cite{Saintmichel2016}. Our ultrasonic imaging technique relies on a custom-made high-frequency scanner driving a vertical array of 128 piezoelectric transducers that send short plane pulses across the gap of the Taylor-Couette cell. The full specifications of this setup can be found in Ref.~\cite{Gallot2013}. In short, \th{the plane ultrasonic pulses are scattered during their propagation either by the sample microstructure or by tracer particles added to the system. The scattered echoes constitute an ultrasonic ``speckle'' that is further processed into images of the echogeneous structures. Cross-correlating successive images yields} the time-resolved local tangential displacement of the sample $\Delta\usv(r, z, t)$ at time $t$ and position $(r,z)$ in cylindrical coordinates (with $r$ the distance to the inner cylinder and $z$ the position along the rotation axis, see sketch in Fig.~\ref{fig:prise_fsweep}). 

In the case of oscillatory shear, LORE uses $\Delta\usv(r, z, t)$ to compute the local strain and the local viscoelastic moduli~\cite{Saintmichel2016}. As shall be checked below, the displacement field $\Delta\usv(r, z, t)$ is most often invariant along $z$ so that we can compute the local strain $\gamma\usv(r,t)$ averaged over the $z$-direction, which dramatically increases the signal-to-noise ratio. Similarly to $\gamma\rheo(t)$, $\gamma\usv(r,t)$ can be further expanded in Fourier series:
\begin{equation}
	\gamma\usv(r,t) = \sum_k \gamma_k\usv(r) \cos \left[ 2 k \pi f t + \phi_k\usv (r) \right]  \,,
\label{eq:gamma_usv_def2}
\end{equation}
where $\gamma_k\usv$ and $\phi_k\usv$ are respectively the amplitude and phase lag of mode $k$. \th{Moreover, the momentum-conservation equation provides the local stress as a function of $r$:}
\begin{equation}
\sigma\rheo(r,t) = \sigma\rheo_1(r) \cos(2 \pi f t) \,,
\label{eq:strain_rheo_def2}
\end{equation}
\th{where the local amplitude $\sigma\rheo_1(r)=\Gamma_1/[2 \pi H (R_{\rm i} + r)^2]$ is deduced from the amplitude $\Gamma_1$ of  the torque $\Gamma(t)$ imposed by the rheometer. Finally, the local viscoelastic moduli $G'_{\rm loc}(r)$ and $G''_{\rm loc}(r)$ are given by:}
\begin{align}
G'_{\rm loc}(r)  &= \frac{\sigma\rheo_1(r)}{\gamma_1\usv(r)} \cos(\phi_1\usv(r))\,, \\
G''_{\rm loc}(r) &= \frac{\sigma\rheo_1(r)}{\gamma_1\usv(r)} \sin(\phi_1\usv(r))\,.
\label{eq:G_usv}
\end{align}

\subsection{Casein gels}
\label{sec:gel}

\begin{figure}
	\centering
  \includegraphics{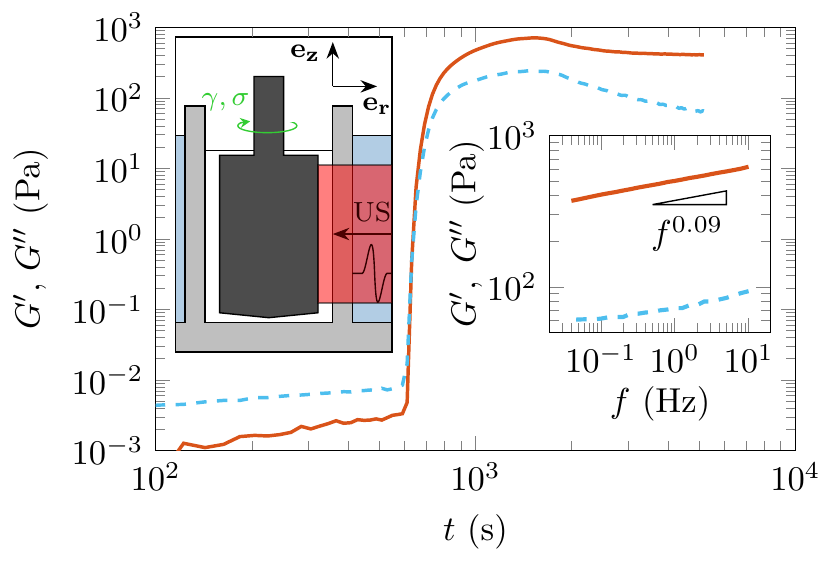}
     \caption{Evolution of the storage modulus $G'$ (orange, solid line) and loss modulus $G''$ (blue, dashed line) during the gelation of a 6\%~wt. casein gel. GDL is added to the aqueous sodium caseinate dispersion at $t=0$~s. Right inset: $G'$ and $G''$ vs. frequency at $t \approx 5000$~s after gelation. Measurements performed under small strain oscillations of amplitude $\gamma\rheo = 0.01$. Left inset: sketch of the LORE experiment. ``US'' and the sine wave denote the ultrasonic plane wave that allows us to measure the local displacements of the gel under oscillatory shear.}
    \label{fig:prise_fsweep}
\end{figure}

Our protein gels are obtained through slow acidification of a dispersion of sodium caseinate (Firmenich) at a weight concentration $[cas]$ in deionized water by using glucono-$\delta$-lactone (GDL, Firmenich) at a weight concentration $[gdl]=[cas]$ ranging from 3.5 to 9\%~wt. In order to provide acoustic contrast and enable LORE measurements, the sodium caseinate dispersion is seeded with neutrally buoyant polyamide spheres (Orgasol 2002 ES3 NAT3, diameter 30~$\mu$m, density $d = 1.03$, Arkema) as described in Ref.~\cite{Leocmach2014}. Immediately after adding  GDL to the sodium caseinate dispersion, the liquid mixture is poured into the Taylor-Couette cell, which defines the initial time $t=0$. GDL then slowly lowers the pH of the aqueous caseinate dispersion over time and eventually induces gelation~\cite{Roefs:1990a,Roefs:1990b}. Casein gels are known to stick to PMMA so that wall slip or syneresis is not an issue in the present geometry~\cite{Leocmach2014}.

Small-amplitude oscillations with a strain amplitude of 1\% are applied to monitor the storage modulus $G'$ and loss modulus $G''$ starting at $t=0$. For a $[cas]=6$\%~wt system, $G'$ overcomes $G''$ after about 600~s signaling gelation (see Fig.~\ref{fig:prise_fsweep}). Around 1500~s, $G'$ and $G''$ reach a maximum then slowly decay to reach a steady value for $t\gtrsim 4000$~s. The slow decrease of $G'$ and $G''$ is attributed to the over-acidification by GDL of sodium caseinate~\cite{Bremer1989,Roefs:1990a,Lucey1997,Leocmach2015}. The resulting gel can be considered as a homogeneous soft solid. Its frequency spectrum is reminiscent of a critical gel:\cite{Winter1986,Jaishankar2012} both $G'$ and $G''$ follow a power law, $G'\sim G''\sim f^{\alpha}$, with $\alpha=0.09$ (see inset of Fig.~\ref{fig:prise_fsweep}). Its structure is well \th{characterized by a fractal network of dimension $\approx 2.1$ and largest cluster size $\approx 4~\mu$m~\cite{Leocmach2015,Lucey1997,Bremer1989,Roefs1986}.} Similar results are obtained for all concentrations investigated here. In all cases, ``instantaneous'' and ``fatigue'' oscillatory shear experiments coupled to LORE are performed in the steady state after letting the gel set for at least 4000~s. 

\begin{figure}
	\centering
    \includegraphics{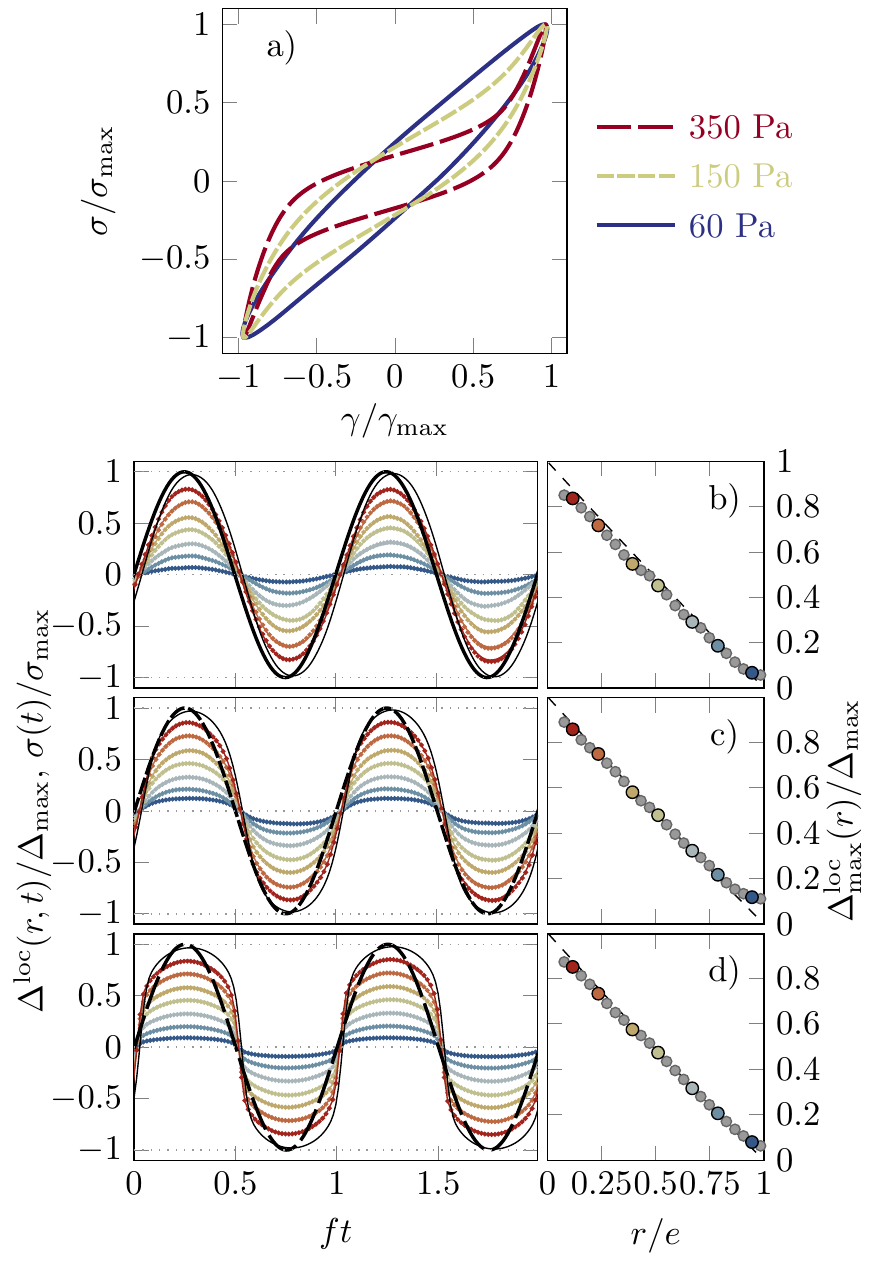}
    \caption{Casein gel at  5.5\% wt. subject to ``instantaneous' oscillatory shear experiments at $f=0.1$~Hz. (a) Lissajous plots of the stress $\sigma\rheo$ vs. strain $\gamma \rheo$ for three different stress amplitudes $\sigma_1=60$~Pa (blue), 150~Pa (yellow) and 350~Pa (red) normalized by the maximum strains $\gamma_\text{max}$ and stresses $\sigma_\text{max}=\sigma_1$. (b,c,d)~Corresponding LORE measurements. (Left panels) Local displacement $\Delta\usv(r,t)$ averaged over the vertical direction for various radial positions $r$ across the gap of the Taylor-Couette cell and normalized by the maximum value in time $\Delta_\text{max}$. The dashed and solid black lines show the normalized applied stress $\sigma\rheo(t)/\sigma_\text{max}$ and the normalized strain response $\gamma\rheo(t)/\gamma_\text{max}$ respectively. (Right panels) Displacement profile $\Delta\usv_\text{max}(r)$ as a function of the normalized radial position $r/e$. The colored dots indicate the radial positions $r$ used for the time series shown in the left panels. The dashed line corresponds to the linear profile expected for a homogeneously strained material.}
    \label{fig:yaourt_response}
\end{figure}

\section{Results and discussion}\label{sec:res}

\subsection{Predicting rupture from ``instantaneous'' oscillatory shear experiments}
\label{sec:instantaneous}

Figure~\ref{fig:yaourt_response} presents the results of  ``instantaneous'' oscillatory shear experiments for $[cas]$=5.5\%~wt. and $f=0.1$~Hz. At low stress amplitude, in the linear viscoelastic regime (LVE), the gel behaves like a linear homogeneous soft solid with $G'= 440$~Pa and $G''= 110$~Pa. The Lissajous-Bowditch representation $\gamma\rheo$ vs $\sigma\rheo$ is elliptical and the strain response is sinusoidal both in rheological and in LORE measurements (see blue curves in Fig.~\ref{fig:yaourt_response}). As the stress amplitude  $\sigma_1$ is increased, the Lissajous-Bowditch representation deviates from an ellipse and displays a distorted shape. Significant harmonic content is evident from the strain and local displacement data (see yellow curves in Fig.~\ref{fig:yaourt_response}). Such a response is typical of nonlinear viscoelasticity (NLVE)~\cite{Hyun2011,Ewoldt2008}. Nonlinearity keeps increasing with the stress amplitude (see red curves in Fig.~\ref{fig:yaourt_response}). Quite remarkably, the local strain field is homogeneous whatever the value of $\sigma_1$, as seen from the linear displacement profiles in Fig.~\ref{fig:yaourt_response}(b-d). Therefore, the gel remains spatially homogeneous and sticks to the walls up to rupture (see also Appendix~A for local measurements of $G'$ and $G''$ and Fourier analysis of $\gamma\usv$). This implies that global rheological data fully reflect the local mechanical properties of the gel. Eventually, the gel experiences a sudden rupture within less than 10 oscillations above some critical oscillatory stress (here $\sigma_c \gtrsim 350$~Pa) corresponding to a critical strain amplitude $\gamma_c$. The local rupture scenario will be described in details in Sect.~\ref{sec:rupture} below.

\begin{figure*}
	\centering
   \includegraphics{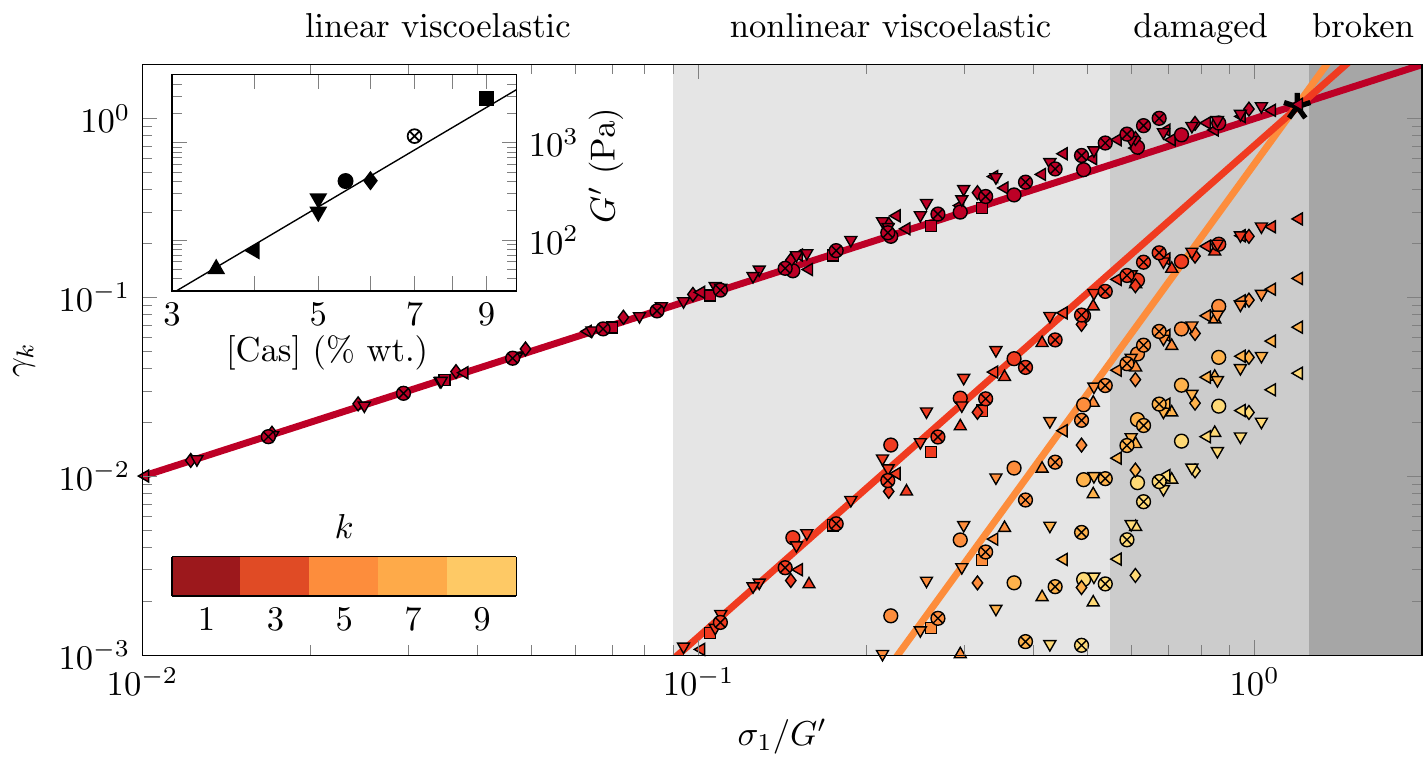}
    \caption{Build-up of nonlinear modes $\gamma\rheo_k$ of the strain response of casein gels subject to an oscillatory stress of amplitude $\sigma_1$ for various casein concentrations. Colors code for $k=1$ to 9. The stress amplitude is normalized by the elastic modulus $G'$ of the gel measured in the LVE regime. Solid lines show the best power-law fits of $\gamma\rheo_k$ for $\sigma_1/G'<0.5$. Background colors delimit the various regimes discussed in the text. Inset: $G'$ as a function of $[cas]$. The symbols code for casein concentration in the main graph. The solid line is $G'\sim [cas]^4$. See Table~\ref{tab:gel} for more details. \th{The coefficients $\gamma\rheo_k$ are only shown down to $10^{-3}$ which corresponds to the maximum amount of nonlinearity observed in the stress input at the largest amplitudes. This conservative threshold ensures that the data can be interpreted as the response to a strictly sinusoidal input \cite{Saintmichel2016}.}}
    \label{fig:yaourt_response2}
\end{figure*}

  The above general picture holds for all gel concentrations under study. Deeper insight into nonlinear behavior is gained by focusing on the harmonic modes $\gamma_k\rheo$ of $\gamma\rheo(t)$ as defined in Eq.~\eqref{eq:gamma_rheo_def2} --see also Appendix~B for more indicators specific of the intracycle oscillatory response. Interestingly, the various Fourier coefficients all collapse onto master curves when plotting $\gamma_k\rheo$ vs $\sigma_1/G'$, where $G'$ is the elastic modulus measured in the LVE regime (see Fig.~\ref{fig:yaourt_response2}), even though  $G'$ increases steeply with the caseinate concentration ($G'\sim [cas]^4$, see inset of Fig.~\ref{fig:yaourt_response2}). Such a rescaling applies to both frequencies investigated in this work, namely $f=0.3$ and 0.1~Hz (see Table~\ref{tab:gel}). In the LVE regime, for $\sigma_1/G'\lesssim 0.1$, one has $\gamma_1\rheo=\sigma_1/G'$ as expected from the definition of $G'$ and the coefficients $\gamma_{k>1}\rheo$ \th{all remain smaller than $\gamma_1/100$}. For $\sigma_1/G'\gtrsim 0.1$, Figure~\ref{fig:yaourt_response2} allows us to distinguish between two different behaviors as nonlinearity increases.

\begin{table*}
\centering
\begin{tabularx}{.68\textwidth}{c c c c c c c c c c}
\hline
\rule{0pt}{12pt}[cas] & 3.5 & 4 & 5 & 5 & 5.5 & 6 & 7 & 9 & All data \\ \hline
\rule{0pt}{13pt}
$G'$ (Pa) 	    & 50   	    & 78   & 230  & 310  & 440 	     & 410  & 1160 & 2840 &  \\ 
$f$ (Hz)		& 	0.1	&	0.1	&	0.3	&	0.3	&	0.1	&	0.1				&	0.3	&	0.1	& 	 \\ \hline
\multicolumn{10}{c}{\rule{0pt}{13pt} (a) Experimental determination of the rupture point}  \\
\rule{0pt}{13pt}
    	    	$\sigma_c$ (Pa) & {42}   & 109  & 259  & 256  & {350}  & 499  & 850  & n/a   &  \\ 
				$\sigma_c/G'$   & {0.84} & 1.40 & 1.12 & 0.83 & {0.79} & 1.21 & 0.73 & n/a   & 1.06$\pm$0.25 \\
				$\gamma_c$      & {0.86} & 1.34 & 1.40 & 1.34 & {1.34} & 1.34 & 1.21 & n/a    & 1.32$\pm$0.20 \\
                \hline   
\multicolumn{10}{c}{\rule{0pt}{13pt}(b) Fit results}  \\
\rule{0pt}{13pt}$A_3$    & 0.8 & 0.7 & 1.2 & 0.7 & 0.6 & 0.5 & 0.5 & 0.4 & 0.6$\pm$ 0.05 \\   
    			$A_5$    & 0.6 & 0.5 & 0.9 & 0.7 & 0.4 & 0.3 & 0.4 & 0.2 & 0.4$\pm$ 0.05 \\  
				$n_3$    & 3.1 & 2.9 & 3.0 & 2.7 & 2.5 & 2.7 & 2.6 & 2.5 & 2.7 $\pm$ 0.1\\   
    			$n_5$    & 4.7 & 4.3 & 4.4 & 4.5 & 3.7 & 4.1 & 4.2 & 3.7 & 4.1 $\pm$ 0.4\\ 
\hline
\multicolumn{10}{c}{\rule{0pt}{13pt} (c) Intersection point using the fits} \\
\rule{0pt}{13pt} 
$\sigma_c/G'$  & 1.14 & 1.26 & 1.04 & 1.11 & 1.35 & 1.49 & 1.30 & 1.78 & 1.33 $\pm$ 0.24\\           
$\gamma_c$     & 1.21 & 1.37 & 1.28 & 0.93 & 1.26 & 1.45 & 1.11 & 1.72 & 1.37 $\pm$ 0.23 \\   
\hline     
\end{tabularx}
\caption{Gel properties for various casein concentrations. Elastic modulus $G'$ measured in the LVE regime and frequency $f$ used in the subsequent ``instantaneous'' oscillatory shear experiments. (a)~Experimental determination of the rupture point ($\sigma_c$,$\gamma_c$). The gel with $[cas]=9$\%~wt. was too stiff for the rheometer to be able to reach rupture. (b)~Best fit parameters with Eq.~\eqref{eq:gamma_fit} for each concentration. (c)~Intersection point as defined in the text. The column ``All data'' shows the average values in (a) and the fit results when using the entire concentration series in (b) and (c).
\label{tab:gel}}
\end{table*}

First, for $0.1\lesssim \sigma_1/G'\lesssim 0.5$, $\gamma_1$ keeps following the same linear law while higher harmonic modes are now clearly measurable and nicely described by power laws of $\sigma_1/G'$. More quantitatively, fitting $\gamma\rheo_3$ and $\gamma\rheo_5$ by power laws of $\sigma_1/G'$ along this NLVE regime,
\begin{align}
	\gamma_k\rheo=A_k \left ( \frac{\sigma_1}{G'}\right ) ^{n_k}\,,
\label{eq:gamma_fit}
\end{align}
yields the exponents $n_3=2.7\pm 0.1$ and $n_5=4.1\pm 0.6$ with prefactors $A_3=0.6$ and $A_5=0.4$. Most strikingly, the extrapolation of both these power-law fits and of the linear regime $\gamma_1\rheo=\sigma_1/G'$ to higher stress amplitudes all intersect at a single point. As shown in Table~\ref{tab:gel}(b) and (c) respectively, this point corresponds to ($\sigma_c/G' = 1.33 \pm 0.24, \gamma_c = 1.37 \pm 0.23$) and matches remarkably well the experimental rupture point ($\sigma_c/G' = 1.06 \pm 0.25, \gamma_c= 1.32 \pm 0.20$). This empirical analysis of the nonlinear strain response thus yields a criterion to \textit{predict} the rupture point based on oscillatory shear experiments at intermediate stresses way before the gels fails. This constitutes our first important result.


\begin{figure}[b!]
	\centering
	\includegraphics{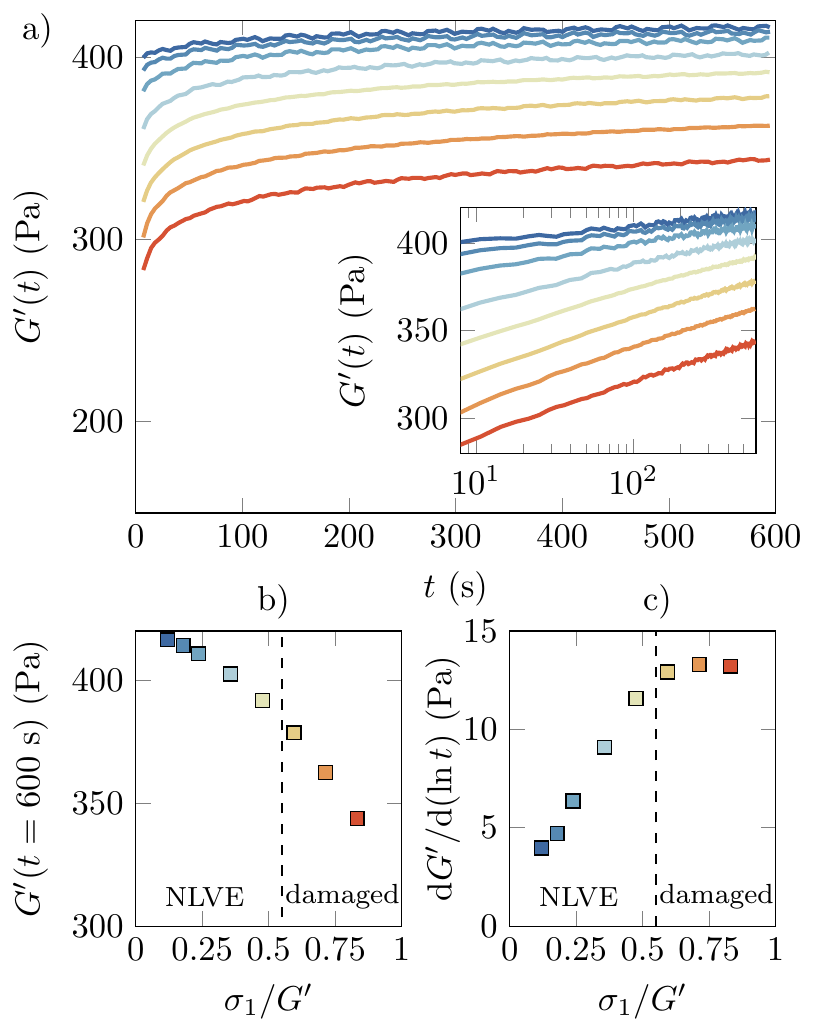}
    \caption{Recovery measurements for a casein gel at $[cas]=5.5$\%~wt. (a)~Elastic modulus $G'(t)$ as a function of time measured under a small oscillatory strain of amplitude $\gamma = 0.01$ right after ``instantaneous'' oscillatory experiments with stress amplitudes $\sigma_1=50$ to 350~Pa (color coded from blue to red). Inset: same data in semi-logarithmic scales. (b)~$G'(t=600~{\rm s})$ as a function of $\sigma_1/G'$ with $G'=440$~Pa the elastic modulus of the intial, undamaged gel. (c)~Slope of the logarithmic increase of $G'$ with $t$. The dashed line delimits the NLVE regime from the damaged regime.}
	\label{fig:recovery}
\end{figure}

Second, for $\sigma_1/G'\gtrsim 0.5$, there is a clear change in the scaling behavior of the higher harmonic modes $\gamma_{k>1}\rheo$ that increase much less steeply with the stress amplitude. Moreover, the fact that $\gamma_1\rheo>\sigma_1/G'$ is indicative of a global softening of the gel. We attribute this change of nonlinear behavior at large stress amplitudes to permanent damage within the gel structure. Such an interpretation is supported by the recovery measurements displayed in Fig.~\ref{fig:recovery}(a) for $[cas]=5.5$\%~wt. There, the time evolution of $G'(t)$ is recorded immediately after an ``instantaneous'' oscillatory stress experiment at amplitude $\sigma_1$ is stopped at time $t=0$. We observe that $G'(t=0)$ lies always below the value of the elastic modulus of the gel at rest, $G'=420$~Pa. However, for $\sigma_1/ G'\lesssim 0.5$, the gel recovers its original elastic modulus within less than 10\% in 10 min [see Fig.~\ref{fig:recovery}(b)]. Moreover, the recovery of $G'(t)$ is characterized by a logarithmic growth with time [see inset of Fig.~\ref{fig:recovery}(a)]. The logarithmic slope increases with $\sigma_1/G'$ for imposed stresses up to $\sigma_1 / G' \simeq 0.5$ above which it remains constant [see Fig.~\ref{fig:recovery}(c)]. We take this as an indication that the gel no longer recovers and remains significantly and irreversibly damaged over the accessible time window for 
$\sigma_1 / G' \gtrsim 0.5$, which nicely corresponds to the change of nonlinear behavior observed in Fig.~\ref{fig:yaourt_response2}. \th{We conclude that over the fitting interval chosen to model $\gamma\rheo_3$ and $\gamma\rheo_5$ in the NLVE regime, the gels remain kinetically reversible and essentially undamaged by the ``instantaneous'' oscillatory shear experiments. Therefore, the fact that power-law extrapolations of $\gamma_{k}\rheo$ cross at the rupture point suggests that the information on gel rupture is essentially embedded in its NLVE regime.}
 
\subsection{Mapping ``fatigue'' experiments onto ``instantaneous'' oscillatory shear measurements}
\label{sec:fatigue}

\begin{figure}[b!]
	\centering
	\includegraphics{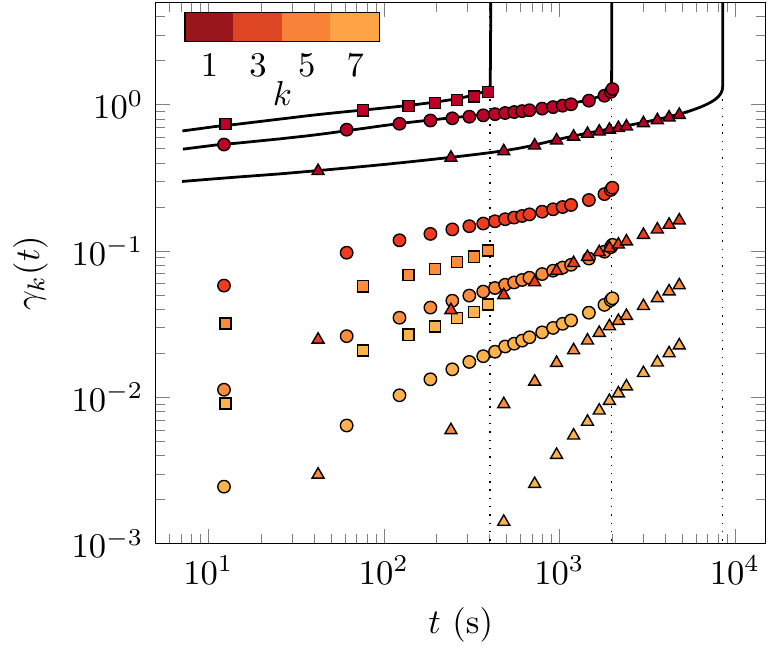}
    \caption{``Fatigue'' experiments on casein gels at $[cas]= 6$\%~wt. An oscillatory stress is applied on a fresh gel at $t=0$ with a constant stress amplitude \th{$\sigma_1\rheo = 90$ ({\small $\triangle$}), 150 ({\Large $\circ$}) and 220~Pa ({\footnotesize $\square$}). These gels respectively break at $\gamma_1 = 1.43$, 1.33 and 1.36}. Black lines represent the strain amplitude $\gamma_\text{max}\rheo$ over time while symbols show the evolution of the harmonic modes $\gamma_k\rheo$. Dotted lines correspond to the rupture times $\tau_f$ where $\gamma_\text{max}\rheo$ diverges.}
	\label{fig:creep}
\end{figure}

\th{The previous ``instantaneous'' oscillatory shear experiments provide an empirical criterion that allows one to predict rupture from the NLVE regime. These experiments were designed to avoid any time-dependence of the gel properties. However, in practice, when the gel is submitted to an oscillatory stress with a large enough constant amplitude, it may also fail over time due to damage accumulation, a phenomenon known as fatigue~\cite{Suresh:1998,Kun2007,Kun2008,Miksic2011,Gibaud2016}. Therefore we now turn our focus on the effect of time for a fixed stress amplitude and ask whether scalings similar to those found above also hold in ``fatigue'' experiments}.

Figure~\ref{fig:creep} shows the strain amplitude $\gamma_\text{max}\rheo$ as a function of time together with the Fourier modes $\gamma_k\rheo(t)$ computed at selected times following the ``fatigue'' protocol described in Sect.~\ref{sec:protocol}. $\gamma_\text{max}\rheo\simeq\gamma_1\rheo$ shows a primary regime characterized by a weak power-law increase with time. The strain then accelerates dramatically up to the sudden rupture of the gel. The rupture time $\tau_f$ sharply decreases for increasing applied stress amplitude. These results are qualitatively very similar to those reported in Ref.~\cite{Leocmach2014} for similar casein gels under a constant load, i.e. for creep experiments (see also Appendix~C). Yet, the present fatigue experiments provide additional access to the progressive build-up of nonlinearity through $\gamma\rheo_{k>1}(t)$ (see symbols in Fig.~\ref{fig:creep}). 

\begin{figure}
	\centering
	\includegraphics{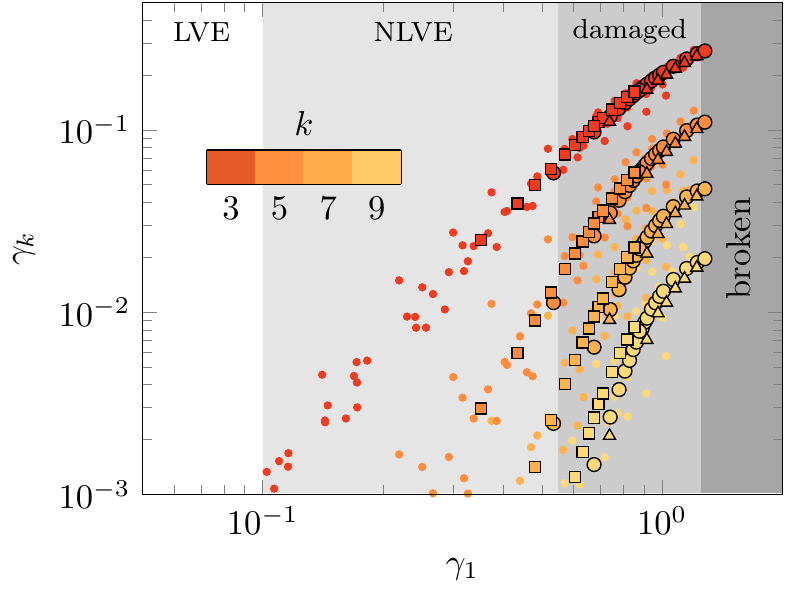}
    \caption{Comparison between ``fatigue'' and ``instantaneous'' experiments. The higher harmonic modes $\gamma\rheo_{k>1}$ are plotted against $\gamma\rheo_1$ for both the ``fatigue'' data in Fig.~\ref{fig:creep} (large symbols) and the ``instantaneous'' data of Fig.~\ref{fig:yaourt_response2} (small dots).}
	\label{fig:creeprescaled}
\end{figure}

In order to compare ``fatigue'' experiments and the previous ``instantaneous'' oscillatory stress experiments, the coefficients $\gamma\rheo_{k>1}$ are plotted as a function of $\gamma\rheo_1$ in Fig.~\ref{fig:creeprescaled}. In this representation, the fatigue data scale perfectly well on top of each other in spite of widely different temporal responses. Moreover, they nicely match the previous $\gamma\rheo_{k>1}$ vs $\gamma\rheo_1$ data from the ``instantaneous'' experiments (see dots in Fig.~\ref{fig:creeprescaled}). We conclude that in both types of oscillatory stress experiments, the same physics govern the build-up of nonlinearity, spanning all three LVE, NLVE and damaged regimes up to rupture. \th{This constitutes our second key result. More precisely, the fact that the dependence on time can be excluded suggests that the whole rupture process is independent of the temporal pathway to failure and only depends on the strain. Further experimental confirmation could be gained by investigating more complex oscillation protocols, e.g. by temporarily reducing the oscillation amplitude in the ``damaged'' regime in order to let the sample relax and check for irreversible evolution. Indeed, while viscous flow through the porous gel structure and/or small plastic deformation should account for losses in the LVE and NLVE regime~\cite{Leocmach2014}, one expects microcrack nucleation and growth to occur under larger stress oscillations in the ``damaged'' regime \cite{Pradhan2010}. Here, unlike creep under a constant stress, we do not observe any obvious cracks or fractures neither through LORE measurements nor through direct observation of the gel surface: the present gels seem to fail through irreversible bond breaking all across the sample. This does not mean that there are no failure precursors or that microcracks below the spatial resolution of the LORE technique are not present in the bulk sample. Still, if macroscopic fractures grow and extend across the whole gap of the Taylor-Couette cell, they do so within a couple of oscillations. This could correspond to a regime of ``diffusive damage'' rather than ``single-crack propagation,'' as recently predicted by a fiber-bundle model \cite{Halasz2012}. Such a picture is supported by the following experiment showing the macroscopic separation of the system into a fluidlike phase and a solidlike phase after rupture.}

\subsection{Monitoring rupture and beyond through LORE}
\label{sec:rupture}
\begin{figure*}
	\centering
    \includegraphics{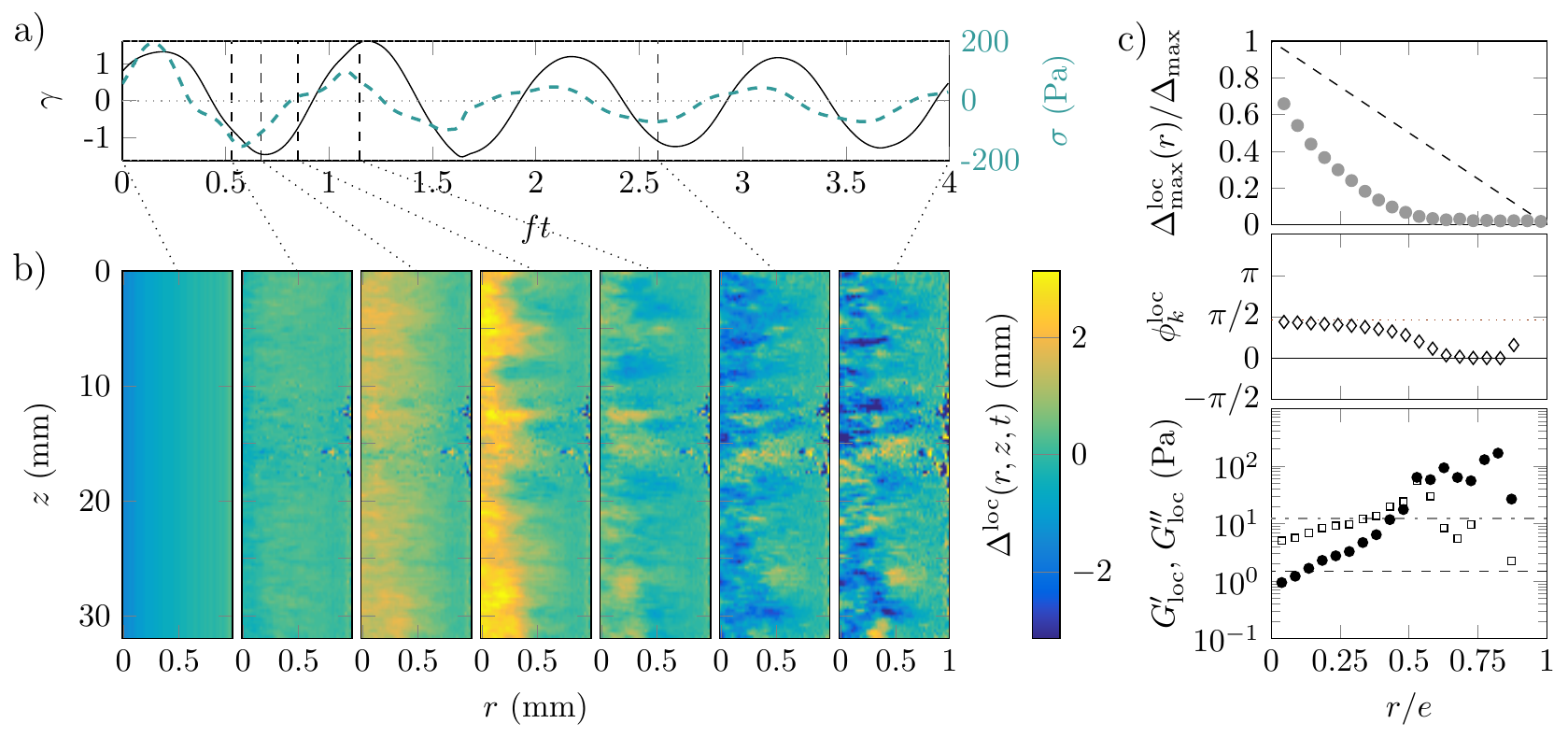}
      \caption{Rupture of a 6\%~wt. casein gel under oscillatory shear. (a)~Strain $\gamma \rheo$ and stress $\sigma \rheo$ as a function of $ft$. Here, an oscillatory strain at $f=0.1$~Hz is applied with amplitude $\gamma\rheo = 1.2$ to prevent the strain from diverging at rupture. Note that the rheometer fails to apply a perfectly constant amplitude $\gamma \rheo$. (b)~Displacement maps $\Delta\usv(r,z,t)$ obtained with LORE during gel breakdown. See also Supplemental Movie~1. (c)~Local measurements in the broken gel at $\sigma_1= 30$~Pa. From top to bottom: $\Delta\usv_\text{max}(r)$, $\phi_1\usv(r)$ and the local viscoelastic moduli $G'_\text{loc}(r)$ ({\large $\bullet$}) and $G''_\text{loc}(r)$ ({\footnotesize $\square$}) as a function of $r/e$. The dashed lines show the theoretical profile for a homogeneous strain field in the Taylor-Couette geometry (top panel) and the phase $\phi_1\rheo$ measured by the rheometer (middle panel). In the bottom panel, the dashed and dashed-dotted lines respectively show $G'$ and $G''$ indicated by the rheometer.}
    \label{fig:break}
\end{figure*}
As noted above for both ``instantaneous'' and ``fatigue'' experiments, the gel breaks irreversibly when the strain $\gamma_1\rheo$ exceeds $\gamma_c\simeq 1.3$. LORE measurements allow us to monitor the rupture process as reported in Fig.~\ref{fig:break} and in Supplemental Movie~1. There, just before rupture, the oscillations are switched from stress-imposed to strain-imposed in order to prevent $\gamma(t)$ from diverging. Time-resolved displacement maps $\Delta\usv(r,z,t)$ show that during gel breakdown, the strain response evolves from perfectly homogeneous to locally highly heterogeneous within a fraction of an oscillation [see Fig.~\ref{fig:break}(b)]. At a given time, $\Delta\usv$ even changes sign from one point to the other. At this stage, the gel is probably broken down into large pieces that may undergo three-dimensional, rotational motion in the oscillatory flow~\cite{Gibaud2016}. 

In order to characterize the steady state resulting from failure, we subsequently apply a small oscillatory stress to the broken gel. The results of the $z$-averaged LORE measurements are displayed in Fig.~\ref{fig:break}(c). While global rheological data indicate a fluid response [i.e. $\gamma\rheo(t)$ in phase quadrature with $\sigma\rheo(t)$, see last oscillation in Fig.~\ref{fig:break}(a)], local measurements actually show that the system has become heterogeneous. This in stark contrast with the initial, homogeneous unbroken gel (see Fig.~\ref{fig:yaourt_response} and Appendix~A). After failure, the strain is localized over about half the gap close to the moving cylinder, whereas for $r/e\gtrsim 0.5$, the strain is negligible. Concomitantly, the local phase lag is $\phi_1\usv\simeq\pi/2$ for $r/e\gtrsim 0.5$ but it goes to zero elsewhere, up to experimental noise close to the fixed wall. The corresponding local estimates of the viscoelastic moduli confirm that the breakdown of the gel leads to two different phases: in the present case, the gel has kept a solid structure with an elastic modulus $G'_{\rm loc}\simeq 200$~Pa for $r/e\gtrsim 0.5$ while for $r/e\lesssim 0.5$, it has become a viscoelastic fluid with viscous modulus $G''_{\rm loc}\simeq 10$~Pa. We also note the presence of wall slip at the moving wall: the local displacement differs from the imposed strain by about 20\%, which is indicative of some degree of syneresis upon failure.

Therefore, our third and final result is that the rupture of casein gels proceeds abruptly in a heterogeneous and gives way to a fluid--solid phase separation.

\section{Conclusion}

We have used oscillatory shear experiments coupled to LORE, a technique that gives access to the local strain field and local viscoelastic moduli, to probe the nonlinear mechanical behavior of casein gels. We have demonstrated that casein gels of different concentrations show a similar mechanical response. Upon applying stresses of increasing amplitude for only 10 oscillations followed by rest periods of 10~min, the gels first deform reversibly up to a relative stress amplitude $\sigma_1/G'\simeq 0.5$ that corresponds to a strain amplitude $\gamma\simeq 0.5$. The higher harmonic modes $\gamma\rheo_k$ of the strain response to an oscillatory stress can be scaled onto master curves as a function of $\sigma_1/G'$ for all concentrations. We have further shown that $\gamma\rheo_{k>1}$ at intermediate stresses are well fitted by simple power laws of $\sigma_1/G'$ which extrapolations at large stress amplitudes point to the rupture conditions $\gamma_c$ and $\sigma_c$. This representation is thus a powerful tool as it permits to predict the rupture point from oscillatory experiments in the nonlinear viscoelastic regime without significantly damaging the gels. Moreover, ``fatigue'' experiments, where a constant stress amplitude is imposed for a long time, map perfectly well onto the previous point-by-point oscillatory experiments when considering $\gamma\rheo_{k>1}$ as a function of $\gamma\rheo_1$. This proves that the rupture pathway is controlled by strain rather than time. Finally, we have shown that the failure process is fast as it occurs within a fraction of a single oscillation. Thanks to LORE, we have checked that the strain field remains locally homogeneous up to rupture. We have visualized the rupture process and shown that it does not involve fracture nucleation and growth. In the steady state following rupture, local measurements reveal a phase separation between a fluid region and a solid region that coexist within the broken gel. 

To summarize, the present results demonstrate for the first time the possibility to predict experimentally the rupture conditions ($\gamma_c, \sigma_c$) of a protein gel based on simple nonlinear analysis of the oscillatory response and without breaking the gel. We expect this work to trigger numerical modeling and theoretical efforts to explain the power-law evolution of the higher harmonic modes as well as the rupture criterion. We believe those scalings and the rupture prediction to be also relevant for biological networks and biopolymer gels made of, e.g., actin, alginate, gelatin~\cite{Baumberger2006} or agar~\cite{Bossi2016}. Further experiments on these various systems are in line to check for generality. Our work also opens the way to deeper local investigations of the damage process of soft materials under oscillatory shear.

\vspace*{10pt}
\section*{Acknowledgments}
The authors wish to thank T.~Divoux, M.~Leocmach, G.~McKinley and B.~Keshavarz for fruitful discussions and A.~Parker at Firmenich for providing the casein and GDL. This work was funded by the European Research Council under the European Union’s Seventh Framework Programme (FP7/2007-2013)/ERC Grant Agreement No. 258803.

\bibliographystyle{apsrev4-1}
\bibliography{biblio}

\clearpage
\section*{Appendix}

\subsection{Local displacement and viscoelastic moduli during ``instantaneous'' oscillatory experiments}
\label{app:lore}

\begin{sidewaysfigure}
	\vspace*{5cm}
    \hspace*{-.6cm}
    \includegraphics[width=24cm]{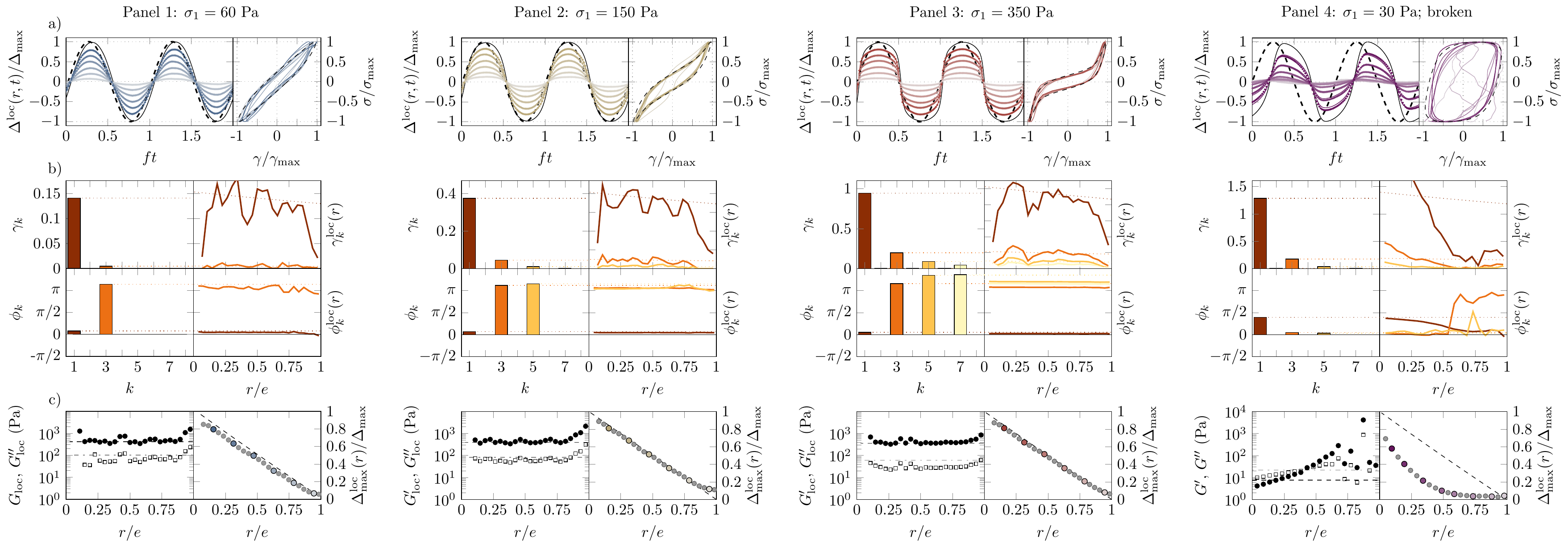}
\caption{``Instantaneous'' oscillatory experiments and LORE on a 5.5\%~wt. casein gel at $\sigma_1=60$~Pa (Panel 1), 150~Pa (Panel 2) and 350~Pa (Panel 3) and on a broken 6\%~wt. casein gel at $\sigma_1 =30$~Pa (Panel 4). The strain amplitudes are respectively $\gamma_{\rm max} = 0.14$, 0.34, 0.82 and 1.24. (a)~Left: local displacements $\Delta\usv(r,t)$ (linear color scale from gray at the stator to colored line at the rotor) and rotor displacement (black solid line) in response to an oscillatory stress $\sigma\rheo(t)$ (black dashed line) as a function of the normalized time $ft$. Right: local Lissajous-Bowditch (from gray to colored lines) and global Lissajous-Bowditch (black dashed line) plots. (b)~Fourier decomposition of the strain response: amplitude $\gamma\rheo_k$ and phase $\phi\rheo_k$ and their local counterparts $\gamma\usv_k$ and $\phi\usv_k$ as a function of $r/e$ (right panels). Colors code for the Fourier mode $k$. (c)~Left: comparison between the local viscoelastic moduli, $G'_{\rm loc}(r)$ ($\bullet$) and $G''_{\rm loc}(r)$ ({\footnotesize $\Box$}) and their rheology counterparts $G'_{\rm rheo}$ (dashed line) and $G''_{\rm rheo}$ (dash-dotted line). Right: maximum normalized local displacement $\Delta_{\rm max}^{\rm loc}(r)/\Delta_{\rm max}$ as a function of the normalized position in the gap $r/e$. The colored dots match the colors and the radial positions $r$ of the lines plotted in subpanel (a). The black dashed line shows the linear, homogeneous profile deduced from rotor displacement.
     }
\label{fig:lore}
\end{sidewaysfigure}

LORE allows us to measure the local mechanical properties of casein gels during ``instantaneous'' oscillatory experiments as explained in Sect.~\ref{sec:mm}. In the LVE regime, $\Delta\usv(r, t)$ remains sinusoidal and in phase with the shear stress $\sigma(t)$ across the entire gap. Its amplitude decreases linearly from the rotor to the stator [see Panel 1(a,b) in Fig.~\ref{fig:lore}] and the local strain at the rotor matches the strain $\gamma(t)$ measured by the rheometer. Correspondingly, in Panel 1(c) of Fig.~\ref{fig:lore}, we observe that the local viscoelastic moduli $G'_{\rm loc}(r)$ and $G''_{\rm loc}(r)$ are constant throughout the gap and perfectly coincide with the rheological measurements:  $G'_{\rm rheo} = 440$~Pa and $G''_{\rm rheo} =80$~Pa for $[cas]=5.5$\%~wt.

In the NLVE and damaged regimes, Panels 2 and 3 of Fig.~\ref{fig:lore} show that the material remains homogeneous and elastic as nonlinearity builds up. The local strain is spatially constant and the fundamental mode is in phase with the applied stress all through the gap. In contrast to the LVE regime, nonlinear modes ($\gamma_k, \phi_k$) emerge for odd values of $k$. It can be checked here that even modes always remain negligible, with at most $\gamma_2\usv / \gamma_1\usv \simeq 0.01$ corresponding to the noise level. Here again, since the protein gel does not slip at the walls and remains spatially homogeneous, the local viscoelastic moduli match the rheological measurements. 

Finally, Panel 4 in Fig.~\ref{fig:lore} gathers the full data corresponding to Fig.~\ref{fig:break}(c) once the gel is broken (see also ESI$^{\ast}$).


\subsection{Intracycle analysis of ``instantaneous'' oscillatory experiments}
\label{app:lissajous}

\begin{figure}
	\centering
    \includegraphics{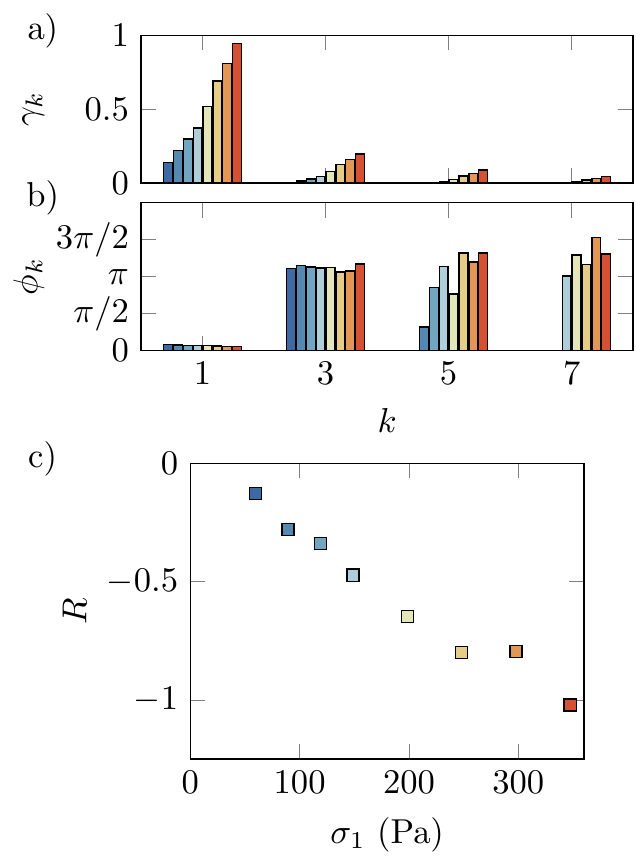}
    \caption{Casein gel at 5.5\%~wt. subject to ``instantaneous'' oscillatory stress experiments. Fourier analysis of the strain response: (a)~amplitude $\gamma_k\rheo$ and (b)~phase $\phi_k\rheo$ as a function of the value of the harmonic number $k$. (c)~Relative ratio of the change in compliance $R$ as a function of $\sigma_1$. Colors code for the imposed stress amplitude from blue (60~Pa) to red (350~Pa).}
    \label{fig:lissajous}
\end{figure}

Classical analysis of oscillatory shear experiments involve Lissajous-Bowditch representation and Fourier transform analysis as discussed in the main text. It was recently proposed in Refs.~\cite{Lauger2010,Dimitriou2012} to analyze the strain response to an imposed sinusoidal stress in terms of Chebyshev polynomials. This approach yields physical insights into the intracycle properties of the gel~\cite{Ewoldt2013,Perge2014}. From the Fourier analysis of Fig.~\ref{fig:lissajous}(a,b), we compute the zero-stress elastic compliance $J'_M$ and the large-stress elastic compliance $J'_L$ defined respectively as:

\begin{alignat}{5}
	J'_M& =\gamma\rheo_1\cos\phi\rheo_1 -3 &&\gamma\rheo_3 \cos\phi\rheo_3 +5 &&\gamma\rheo_5 \cos\phi\rheo_5 -...\\
    J'_L& =\gamma\rheo_1\cos\phi\rheo_1 +  &&\gamma\rheo_3 \cos\phi\rheo_3 +  &&\gamma\rheo_5\cos\phi\rheo_5  +....
\label{eq:lis}
\end{alignat}
The measurements of $J'_M$ and $J'_L$ allow one to calculate the relative ratio of the change in compliance $R$:
\begin{align}
	R& = (J'_L-J'_M)/J'_L.
\label{eq:lis2}
\end{align}

As discussed in Ref.~\cite{Dimitriou2012}, positive values of $R$ correspond to intracycle stress-softening while negative values indicate stress-stiffening. Figure~\ref{fig:lissajous}(c) indicates that our casein gels show more and more intracycle stress-stiffening as the stress amplitude is increased in ``instantaneous'' oscillatory experiments. 

In the damaged regime in Fig.~\ref{fig:yaourt_response2}, we noted that experimental measurements of $\gamma_1$ fell above the straight line $\gamma_1=\sigma_1/G'$, which indicates a global stress-softening. This is however not in contradiction with the present intracycle analysis. Indeed, during one oscillation, the gel tends to resist stress and stiffens ($R<0$) but from one ``instantaneous'' experiment to the other, the gel becomes softer due to damage accumulation.

\subsection{Comparison between creep and fatigue experiments}
\label{app:creep}

\begin{figure} 
	\centering
	\includegraphics{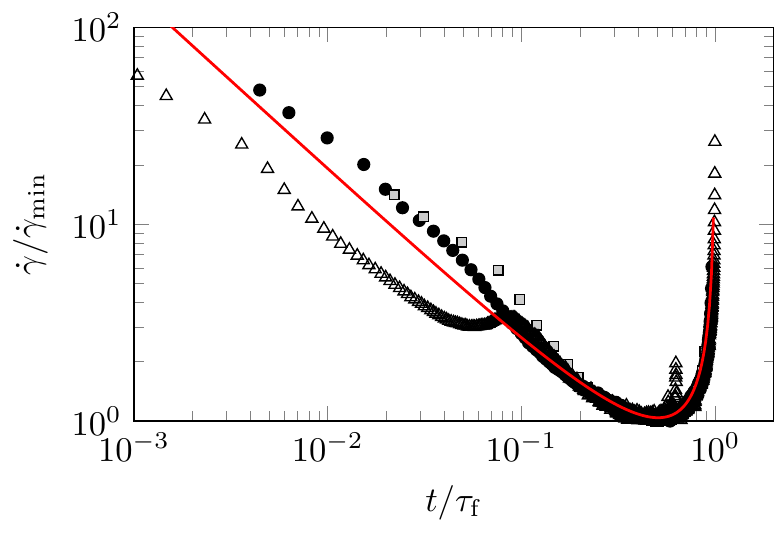}
    \caption{Evolution of the time-derivative of the strain amplitude $\dot\gamma$ normalized by its minimum value $\dot{\gamma}_{\rm min}$ as a function of $t/\tau_\text{f}$ during ``fatigue'' experiments at \th{ $\sigma_1 \rheo$ = 90 ({\small $\triangle$}), 150 ({\Large $\circ$}) and 220 ({\footnotesize $\square$})~Pa}. $\tau_\text{f}$ is the rupture time shown in Fig.~\ref{fig:creep}. The red line corresponds to the best fit using Eq.~\eqref{eq:creep} with $\alpha=0.09$, $\lambda=0.30$ and $\mu=0.24$.}
	\label{fig:ccreep}
\end{figure}

Inspired by previous creep experiments~\cite{Leocmach2014}, Fig.~\ref{fig:ccreep} presents the time derivative of the strain amplitude $\dot{\gamma}={\rm d}\gamma_{\rm max}/{\rm d}t$ normalized by its minimum value $\dot{\gamma}_{\rm min}=\min({\rm d}\gamma_{\rm max}/{\rm d}t)$ for the data shown in Fig.~\ref{fig:creep}. As suggested in Ref.~\cite{Leocmach2014}, plotting $\dot{\gamma}/\dot{\gamma}_{\rm min}$ as a function of $t/\tau_f$ leads to a master curve that is well captured by:
\begin{equation}
	\frac{\dot{\gamma}_{}}{\dot{\gamma}_{\rm min}}=\lambda (t/\tau_f)^{-(1-\alpha)}+\frac{\mu}{1-t/\tau_f}\,
\label{eq:creep}
\end{equation}
where the exponent $\alpha=0.09$ corresponds to that of the frequency-dependence of $G'$ and $G''$ (see inset of Fig.~\ref{fig:prise_fsweep}).

The fatigue experiments at $\sigma_1= 150$ and 220~Pa very nicely follow the scaling of Eq.~\eqref{eq:creep} with an exponent $\alpha$=0.09 for the primary power-law creep regime. This value coincides with the exponent measured in the frequency sweep experiment (see inset of Fig.~\ref{fig:prise_fsweep}), in striking agreement with Ref.~\cite{Leocmach2014}. Thus, in both creep and fatigue experiments, the exponent of the primary regime is governed by linear viscoelasticity. Yet, in the present oscillatory experiments, we do not observe any fracture growth as reported in Ref.~\cite{Leocmach2014}, which suggests that damage induced by fatigue is diffuse rather than localized in fractures. Moreover, at short times, the experiment at $\sigma_1= 90$~Pa does not scale with the experiments under a larger stress amplitude. It rather goes through two maxima in $\dot{\gamma}(t)$, hinting at rupture precursors followed by self-healing. Deeper analysis of rupture times under fatigue and of the shape of $\dot{\gamma}(t)$ is left for future work. 

\clearpage

\clearpage


\end{document}